\documentclass{sigchi}

\CopyrightYear{2019}
\setcopyright{acmlicensed}
\doi{https://doi.org/10.1145/3332165.3347925}
\isbn{978-1-4503-6816-2/19/10}
\conferenceinfo{UIST'19,}{October 20--23, 2019, New Orleans, LA, USA}
\acmPrice{\$15.00}

\usepackage{balance}       % to better equalize the last page
\usepackage{graphics}      % for EPS, load graphicx instead
\usepackage[T1]{fontenc}   % for umlauts and other diaeresis
\usepackage{txfonts}
\usepackage{mathptmx}
\usepackage[pdflang={en-US},pdftex]{hyperref}
\usepackage{color}
\usepackage{booktabs}
\usepackage{textcomp}
\usepackage{tikzpagenodes} % For position hack in Fig 4

\usepackage{microtype}        % Improved Tracking and Kerning
\usepackage{ccicons}          % Cite your images correctly!

\usepackage{todonotes}

\usepackage{wrapfig}
\usepackage{listings}
\usepackage{amssymb}% http://ctan.org/pkg/amssymb
\usepackage{pifont}% http://ctan.org/pkg/pifont for open stars etc

\newcommand{\codeFontSize}
  {\footnotesize}

\newcommand{\levelThree}[1]
  {\paragraph{#1}}
\def\parahead#1
  {\paragraph*{#1}}
\def\paraheadNoDot#1
  {\paragraph*{#1}}
\newcommand{\ie}{{i.e.}}
\newcommand{\eg}{{e.g.}}
\newcommand{\etc}{{etc.}}
\newcommand{\cf}{{cf.}}

\newcommand{\sns}{\ensuremath{\textsc{Sketch-n-Sketch}}}

\newcommand{\wwid}{\emph{Watch What I Do: Programming by Demonstration}}
\newcommand{\wwidPbd}{WWID: PBD}

\newcommand{\suppMaterials}
  {Supplementary Materials}
\newcommand{\autorefFigTwoPaneUI}
  {Figure 1}
\newcommand{\autorefFigExamples}
  {\autoref{fig:examples}}
\newcommand{\citeSnsUist}
  {[19]}
\newcommand{\citeSnsPldi}
  {[9]}
\newcommand{\citeWWIDBenchmarks}
  {[34]}
\newcommand{\citeLillicon}
  {[5]}
\newcommand{\citeQuickDraw}
  {[8]}

\newcommand{\myurl}[1]{\url{#1}}

\newcommand{\codeSize}
  {\small}

\newcommand{\toolName}[1]
  {\textsc{#1}}
\newcommand{\toolNameFirstUse}[1]
  {\toolName{#1}}
\newcommand{\blackStar}{\ding{72}}
\newcommand{\openStar}{\ding{73}}

\newcommand{\hide}[1]{}

\newcommand{\numExamples}{16}
\newcommand{\numLoc}{427}

\newcommand{\miniSepTwo}{\hspace{0.02in}}
\newcommand{\miniSepThree}{\hspace{0.03in}}

\newcommand{\expStr}[1]
  {\ensuremath{\texttt{'{#1}'}}}

\newcommand{\figSyntaxEnd}{\end{array}$}

\newcommand{\figSyntaxBegin}{$\begin{array}{rcl}}

\newcommand{\envExp}[2]
  {{#1}\vdash{#2}}

\newcommand{\reducesTo}[3]
  {\envExp{#1}{#2}\hspace{0.01in}\begingroup\color{black}\Downarrow\endgroup{#3}}

\usepackage{enumitem}
\setlist
  { leftmargin = 14pt
  , rightmargin = 6pt
  , itemsep = 1pt
  , topsep = 0pt
  }

\def\plaintitle{Sketch-n-Sketch: Output-Directed Programming for SVG}
\def\notplaintitle{Sketch-n-Sketch: Output-Directed Programming for SVG}
\def\plainauthor{Brian Hempel, Justin Lubin, Ravi Chugh}

\def\plainkeywords{Output-Directed Programming; SVG; Sketch-n-Sketch}

\makeatletter
\def\url@leostyle{%
  \@ifundefined{selectfont}{
    \def\UrlFont{\sf}
  }{
    \def\UrlFont{\small\bf\ttfamily}
  }}
\makeatother
\def\pprw{8.5in}
\def\pprh{11in}

\definecolor{linkColor}{RGB}{6,125,233}
\hypersetup{%
  pdftitle={\plaintitle},
  pdfauthor={\plainauthor},
  pdfkeywords={\plainkeywords},
  pdfdisplaydoctitle=true, % For Accessibility
  bookmarksnumbered,
  pdfstartview={FitH},
  colorlinks,
  citecolor=black,
  filecolor=black,
  linkcolor=black,
  urlcolor=linkColor,
  breaklinks=true,
  hypertexnames=false
}

\begin{document}

\title{\notplaintitle}

% Following https://en.wikibooks.org/wiki/LaTeX/Source_Code_Listings
\definecolor{basicColor}{rgb}{0.2,0.2,0.4}
\definecolor{keywordColor}{rgb}{0.4,0.0,0.4}
\definecolor{commentColor}{rgb}{0.0,0.5,0.0}
\definecolor{identifierColor}{rgb}{0.0,0.0,0.0}
\lstdefinestyle{customElm}{
  backgroundcolor={}, % the lipics template sets an ugly, hard-to-read color for source listings. Ignore it.
%  belowcaptionskip=1\baselineskip,
  columns=[l]fullflexible, % use font-native character spacing
%  basewidth=0.51em, % character spacing
  basicstyle=\small\ttfamily\color{basicColor}, % same as lipics except for color
%  basicstyle=\linespread{1.05}\ttfamily\color{basicColor}, % line spacing and larger font
  breaklines=false, % don't automatically break lines
  keepspaces=true,
  breakatwhitespace=true,
  xleftmargin=\parindent,
  language=Haskell,
  showstringspaces=false,
  morekeywords={as,zeroTo,??terminationCondition},
  keywordstyle=\bfseries\color{keywordColor},
  commentstyle=\color{commentColor},
  identifierstyle=\color{identifierColor},
  stringstyle=\color{orange}
}

\lstset{style=customElm}

\numberofauthors{1}
\author{
  \alignauthor Brian Hempel \hspace{0.35in} Justin Lubin \hspace{0.35in} Ravi Chugh\\[2pt]
    \affaddr{Department of Computer Science, University of Chicago, USA}\\
    \texttt{\{brianhempel,justinlubin,rchugh\}\hspace{0.02in}@\hspace{0.03in}uchicago.edu}
}

\maketitle

\begin{abstract}

For creative tasks, programmers face a choice: Use a GUI and sacrifice
flexibility, or write code and sacrifice ergonomics?

To obtain both flexibility \emph{and} ease of use, a number of systems have
explored a workflow that we call \emph{output-directed programming}.  In this
paradigm, direct manipulation of the program's graphical output corresponds to
writing code in a general-purpose programming language, and edits not possible
with the mouse can still be enacted through ordinary text edits to the program.
Such capabilities provide hope for integrating graphical user interfaces into what
are currently text-centric programming environments.

To further advance this vision, we present a variety of new output-directed
techniques that extend the expressive power of \sns{}, an output-directed
programming system for creating programs that generate vector graphics.
\mbox{To enable} output-directed interaction at more stages of program construction, we
expose intermediate execution products for manipulation and we present a
mechanism for contextual drawing.
Looking forward to output-directed programming beyond vector graphics, we also
offer generic refactorings through the GUI, and our techniques employ a
domain-agnostic provenance tracing scheme.

To demonstrate the improved expressiveness, we implement a dozen new parametric
designs in \sns{} \emph{without} text-based edits.
Among these is the first demonstration of building a recursive function in an
output-directed programming setting.

\end{abstract}

%\category{H.5.2.}{Information Interfaces and Presentation (e.g. HCI)}{User Interfaces}
%\category{D.2.6.}{Software Engineering}{Programming Environments}
%\category{D.3.3.}{Programming Languages}{Language Constructs and Features}

% ACM Classfication




% Print the classification codes

% This needs to be inside \begin{document} to not screw up the columns.

%% \setlength{\columnsep}{7pt}
\setlength{\columnsep}{10pt}
\setlength{\intextsep}{0pt}

\clearpage
\setcounter{figure}{1} %% b/c teaser hack

\section{Introduction}
\label{sec:introduction}

Direct manipulation~\cite{Shneiderman1983} graphical user interface (GUI) applications are ubiquitous. Every day, direct manipulation GUIs are used by millions for office tasks such as presentation and document preparation as well as for specialized creative tasks such as engineering and graphic design. Direct manipulation's intuitive \emph{point-click-modify} experience enables a large number of people to leverage computers to create novel artifacts.

Even so, experts may sometimes forgo a GUI application and create their artifact using a programming language. In contrast to the intuitive experience offered by direct manipulation GUIs, text-based coding does not offer immediate and direct artifact construction, but programmers use text-based languages to gain flexibility not afforded by any GUI application.

Both paradigms---direct manipulation and text-based programming---have proven useful. Naturally, there have been efforts to combine their distinct strengths.

For example, some professional creative applications do offer a scripting API (e.g. Maya~\cite{Maya}), but, beyond perhaps an initial macro recording, script construction is a text editing process with no direct manipulation support. Furthermore, in the programming world, non-text paradigms such as blocks-~\cite{logoblocks} or wires-based visual programming~\cite{theOnLineGraphicalSpecificationOfComputerProcedures} allow direct manipulation of program elements.
Even so, most professional programming remains a strictly plain text activity.

Instead, could direct manipulation \emph{augment} text-based programming, providing user interfaces more like standard creative application GUIs, such as graphics editors? Rather than directly manipulating program elements construed as blocks and wires, could one construct a textual program by directly manipulating a program's \emph{output}?

\parahead{Output-Directed Programming (ODP)}

We refer to this paradigm---in which a programmer constructs a plain text program by mouse-based operations on the program's output---as \emph{output-directed programming (ODP)}. Each operation triggers a transformation on the text-based program, akin to automatic refactoring tools. Successive operations build the program step by step. Because the program is plain text throughout the process, desired changes that are not possible through the provided direct manipulation interactions can be enacted by ordinary text editing. The system treats the plain text as the artifact's primary representation so that text editing does not disable future direct manipulation actions.

This ODP paradigm for altering a program by manipulating its output offers two tantalizing possibilities:

\begin{itemize}
  \item Creative applications could represent the user's artifact not as an opaque internal data structure but as a visible, editable, textual program. Direct manipulation of the artifact can be freely mixed with computational generation.
  \item For general-purpose programming, ODP could facilitate \emph{general} program construction via output-directed manipulations. For experts, these ODP interactions might accelerate common program construction tasks, and for novices, ODP interactions might provide an approachable pathway into programming.
\end{itemize}

A number of systems have made initial steps towards realizing this paradigm. After first drafting a program using ordinary text edits, several ODP systems allow direct manipulation of the output to change constant literals in the program~\cite{WangEtAl,sns-pldi,carbideAlpha,LivePBE,sns-oopsla}. To also relieve the initial text editing burden, for programs with graphical output a few systems provide ODP mechanisms for program construction~\cite{McDirmidAPX, transmorphic, sns-uist}, akin to drawing tools in graphics editors.

All prior ODP systems presented their workflow as a mix of text edits and direct manipulation.
Although the selling point of ODP is that text edits can compensate for any missing direct manipulation features, the standing question is how thoroughly direct manipulation can subsume text-based editing.
Therefore, in this work we wonder: \textbf{What kinds of programs can be constructed \emph{entirely} through output manipulations?}

\parahead{New ODP Techniques in \sns{}}

We extend our prior work on \sns{}~\cite{sns-pldi,sns-uist}, \mbox{a programming} system for creating vector graphics, with new ODP techniques that enable the system to construct \numExamples{} example programs \emph{without} text edits on the code---even though ordinary text editing remains possible at any time during the construction of each example.
Specifically, we:

\begin{enumerate}

\item[1.]
Expose \emph{intermediate execution values} for manipulation, instead of just
the final output.

\item[2.]
Offer \emph{expression focusing} to enable contextual drawing.

\item[3.]
Expose \emph{generic code refactoring tools} through output-directed
interactions.

\item[4.]
Use \emph{generic} run-time tracing to track \emph{value
provenance}, to associate output values with source code locations.

\end{enumerate}

Although the \sns{} system is specialized for programs that output vector graphics, the four principles above are relevant for future output-directed programming systems of all kinds.

\parahead{Paper Outline and Supplementary Materials}

To introduce the output-directed programming workflow in \sns{}, in the next section we walk through the creation of a simple example program. After this overview, we present the core ODP mechanisms and our four innovations in more detail, followed by an examination of the construction of the \numExamples{} example programs. We conclude by discussing current limitations of \sns{}, related work, and future directions for ODP.
Our implementation, examples, videos, and appendices are available as \suppMaterials{} and on the Web \mbox{(\url{http://ravichugh.github.io/sketch-n-sketch/})}.

\section{Overview}
\label{sec:overview}

\sns{} is a bimodal programming environment, as depicted in \autorefFigTwoPaneUI{}: the left pane is an ordinary source code text editor; the right pane renders the scalable vector graphics (SVG)~\cite{SVG} design generated by the code and also offers a graphical interface for performing transformations on the SVG output (\ie{}, with mouse-based manipulations).
The programmer may perform keyboard text edits on the code at any time during program construction, but this ability will not be used in this paper.

\newcommand{\narrowWrapfigCaption}[1]
  {#1}
  %% {\caption{\\{#1}}}
  %% {\caption{\\[1pt]{#1}}}

\begin{wrapfigure}{r}{0pt}
	\includegraphics[width=0.7in]{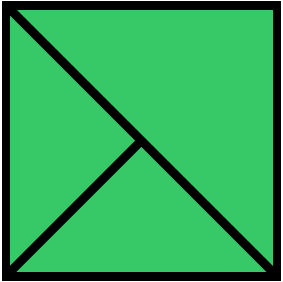}
  \caption{\newline Lambda logo.}
  \label{fig:logoFinal}
\end{wrapfigure}

To introduce the basic ODP workflow in \sns{}, we will walk through the construction of a reusable design for the \sns{} logo (\autoref{fig:logoFinal}).
The construction will proceed in four consecutive phases:
(i) \emph{drawing} the shapes,
(ii) \emph{relating} graphical features to align shapes and equate colors,
(iii) \emph{abstracting} the design into a reusable function, and, finally,
(iv) \emph{refactoring} the design to clean up the generated code.
Throughout, the code shown is as produced by \sns{}.
Newlines inserted for formatting purposes in this paper are ``escaped'' by a ``\verb+\+'' backslash.
The final code for this example is shown in \autoref{fig:logoFinalCode}.

\subsection{Drawing Shapes}
\label{sec:drawingShapes}

The initial program template is nearly blank, defining only an empty list of SVG shapes:
\begin{lstlisting}
svg (concat [
])
\end{lstlisting}

\noindent
Starting from this bare program, the programmer would first like to get some shapes into the program's output. A traditional programming environment would require looking up the drawing library's documentation and perhaps copy-pasting some example code into the program.

The \sns{} output pane, however, imitates a traditional drawing application. The programmer clicks on the ``square'' tool from the toolbox ({\autorefFigTwoPaneUI}a) and drags the mouse on the canvas. When she releases the mouse, a new \verb+square1+ definition is inserted in the program, based on a call to the \verb+square+ standard library function with arguments for position and size derived from the mouse drag operation. A \verb+square1+ variable usage is also inserted into the shape list at the end of the program, and thereby a red square appears in the program output on the canvas (not shown).

\begin{lstlisting}
square1 = square 0 [158, 127] 156

svg (concat [
  [square1]
])
\end{lstlisting}

\begin{wrapfigure}{r}{0pt}
	\includegraphics[width=1.4in]{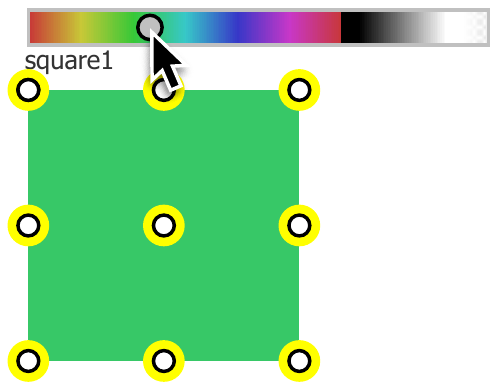}
	\caption{Color slider.}
    \label{fig:logoColorSlider}
\end{wrapfigure}

\begin{figure}[t]
\begin{lstlisting}[numbers=none,xrightmargin=.0\textwidth]
y = 127

x = 158

topLeft = [x, y]

w = 156

square1 = square 140 topLeft w

y2 = y + w

line1 = line 0 5 topLeft [ x+ w, y2]

line2 = line 0 5 [x, y2] \
          [ (2! * x + w)/ 2!, (2! * y + w) / 2!]
...
\end{lstlisting}
\begin{tikzpicture}[remember picture,overlay,shift={(current page.north east)}]
\node[anchor=north east,xshift=-0.65in,yshift=-0.9in]{\includegraphics[width=0.85in]{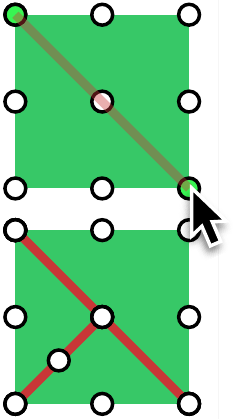}};
\end{tikzpicture}\vspace{-1.1em}
\caption{Code after drawing one square and snap-drawing two lines (right).
\emph{Freeze annotations}, written \texttt{!}, tell \sns{} not to change those constants when shapes are moved on the canvas~\citeSnsPldi.
}
\label{fig:code-after-snap-draw-lines}
\end{figure}

Shapes on the canvas may be directly moved and resized as in a traditional graphics editor, causing \sns{} to automatically change appropriate numeric constants in the program~\cite{sns-pldi}. Additionally, when a shape is selected, sliders appear to allow similar modification of non-shape attributes such as, \eg{}, fill color (\autoref{fig:logoColorSlider}) or stroke width. In our example, the programmer drags the color slider to change the square's fill color from red to a shade of green; the appropriate number in the program is updated (the color number \verb+0+ above becomes \verb+140+, shown in \autoref{fig:code-after-snap-draw-lines}).

To create the lines for the lambda symbol, the programmer selects the ``line'' tool and then ``snap-draws''~\cite{BriarConstraintDrawing} two lines between point features of the green square: she draws one line from the top-left to the bottom-right corner and a second line from the bottom-left corner to the center point (these two steps are shown in the right side of \autoref{fig:code-after-snap-draw-lines}). \sns{} interprets these snaps as constraints which should always hold, \ie{}, the lines should still coincide with the corners and center even if the square is moved or resized.
\sns{} encodes these constraints in the program via shared variables: three new variables are introduced for the square's top-left $x$ and $y$ coordinates and its width $w$, and the spatial properties of both the square and the two lines are defined in terms of simple math on those variables (\autoref{fig:code-after-snap-draw-lines}).

%\pagebreak
\hspace{1em}
\subsection{Relating Properties}
\label{sec:relatingProperties}

\begin{wrapfigure}{r}{0pt}
	\includegraphics[width=2.07in]{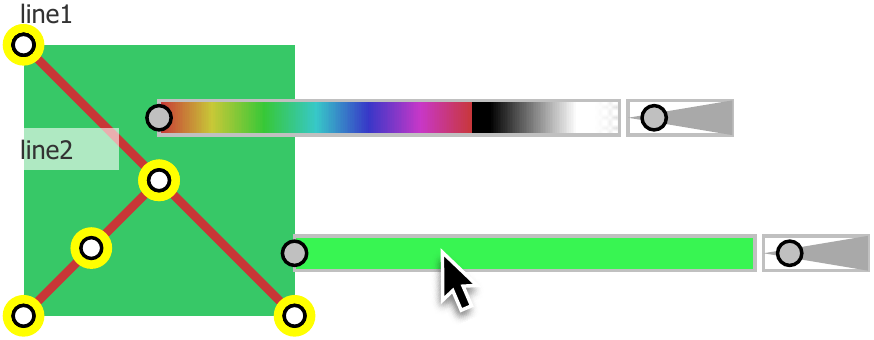}
  \caption{Selecting a feature.}
  \label{fig:logoColorSliders}
\end{wrapfigure}

The programmer would like the color and width of the two lines always to be identical.
\sns{} offers tools for relating features after they are drawn. To relate the colors, the programmer first selects the two lines (indicated by yellow highlights in \autoref{fig:logoColorSliders}) to expose sliders for each line's color and stroke width.
As mentioned earlier, she can manipulate the sliders to change constants in the program, but here she instead clicks the whole \verb+line2+ color slider
to select the property itself (a selected property is indicated in green, as shown in \autoref{fig:logoColorSliders}) and then selects the other line's color slider as well.

\begin{figure}[t]
  \centerline{\includegraphics[width=2.6in]{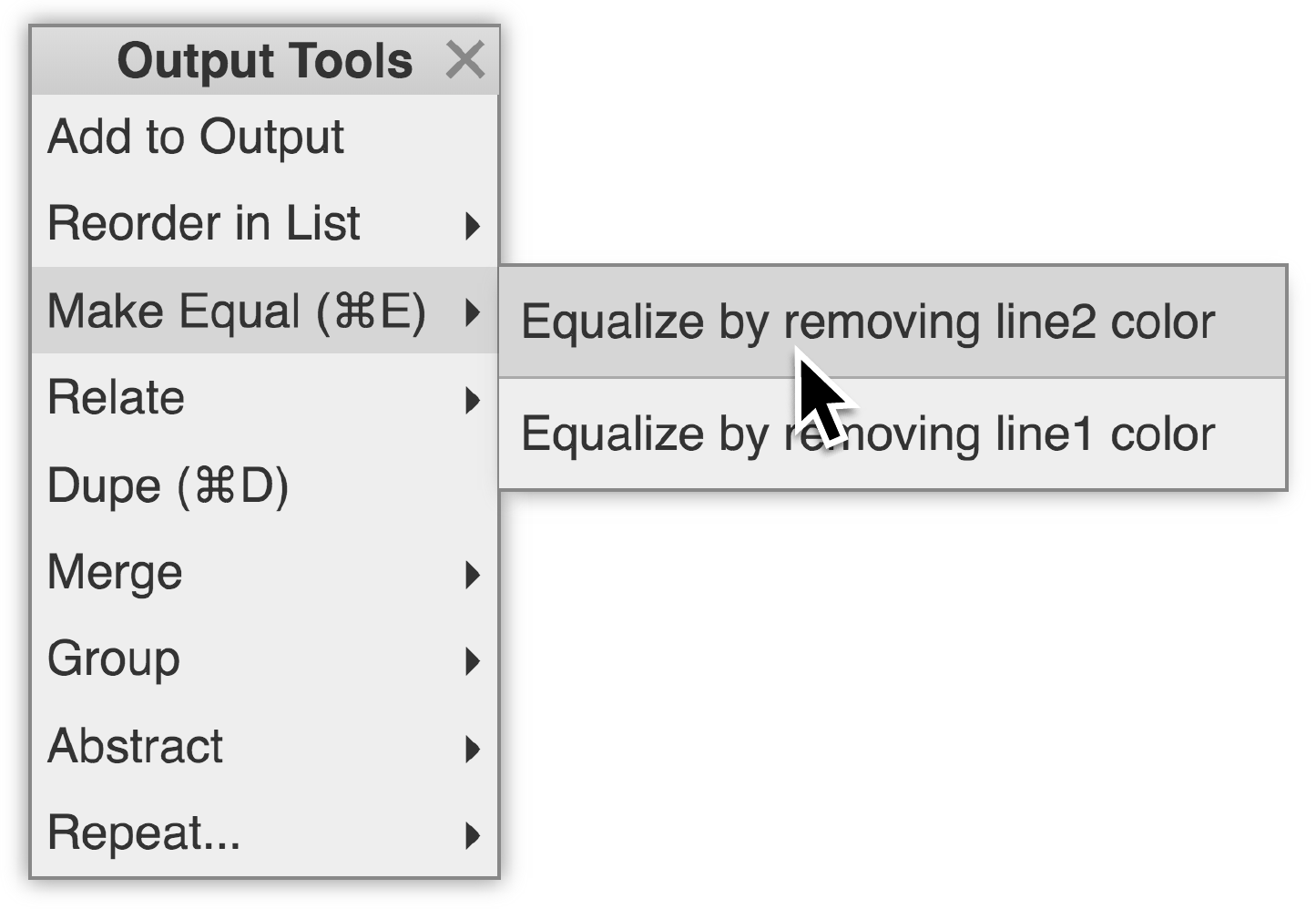}}
  \caption{The output tools panel appears when an item on the canvas is selected. Each tool may offer multiple results; hovering the mouse over each result previews the change on the code and on the canvas.}
  \label{fig:outputToolsPanel}
\end{figure}

Whenever an item is selected, \sns{} displays a floating menu of output tools. The programmer would like the two selected colors to always be the same, so she moves her mouse to the \toolName{Make Equal} tool, revealing a list of possible program transformations in a submenu---in this case there is only one, but if the lines had been different colors two results would be offered to let the programmer decide which line's color should take priority (as simulated in \autoref{fig:outputToolsPanel}). Moving her mouse over the single ``Equalize by removing line2 color'' result in the submenu shows a preview of the new output on the canvas and presents a diff of the change in the code box (not shown)---in this case the appearance will not change but an appropriately-named \verb+color+ variable will be introduced and used for both lines. The programmer is happy with this change and clicks to apply it. She repeats this workflow to equalize the stroke width of the two lines as well.

\subsection{Abstracting the Design}
\label{sec:abstractingTheDesign}

The programmer would like to make her logo design reusable---in other words, she would like to create a \emph{function} that, given size and color arguments, generates a logo appropriately.

She gathers the three shapes into a single expression by selecting the shapes and invoking the \toolName{Group} tool in the floating menu. Analogous to the grouping functionality of traditional graphics editors, \toolName{Group} in \sns{} gathers the shapes into a single list definition, which is used in place of the individual shapes in the final shape list:

\begin{lstlisting}
...
squareLineLine = [square1, line1, line2]

svg (concat [
  squareLineLine
])
\end{lstlisting}

\begin{wrapfigure}{r}{0pt}
	\includegraphics[width=1.5in]{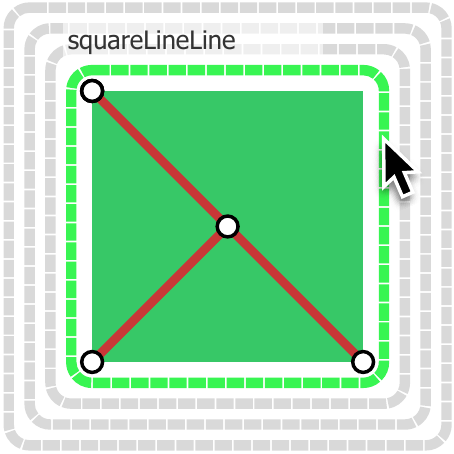}
  \caption{List widgets.}
  \label{fig:logoListWidgetSelected}
\end{wrapfigure}

Now that the shapes are grouped into a list, \sns{} offers that list itself as an object for selection on the canvas. Lists are represented on the canvas with \emph{list widgets}---boxes with dotted borders (intended to evoke the elements of a list). When the programmer hovers her mouse over a shape, list widgets are shown for each of the list expressions in the program in which the shape appeared. In this example there are four such lists: the initial list literal \verb+[square1, line1, line2]+ plus three expressions at the end of the program that evaluate to lists (the \verb+squareLineLine+ variable usage, the list literal \verb+[squareLineLine]+, and the flattened list value produced by the call \verb+concat [squareLineLine]+).
Notice in \autoref{fig:logoListWidgetSelected} how \sns{} chooses a layout to avoid otherwise-overlapping widgets.
Hovering over a list widget highlights the corresponding code expression in the program (not shown), allowing the programmer to distinguish the various lists. She selects the list widget corresponding to the \verb+[square1, line1, line2]+ literal (as shown in \autoref{fig:logoListWidgetSelected}). Because this list literal is assigned to a variable, the list widget is labeled with the variable name, \verb+squareLineLine+.

\begin{wrapfigure}{r}{0pt}
	\includegraphics[width=1.1in]{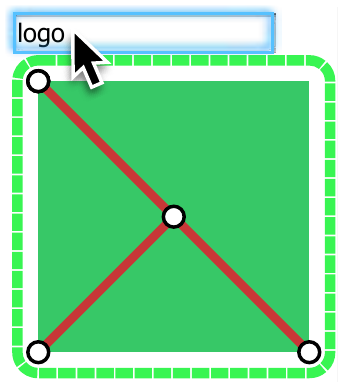}
  \caption{Renaming.}
  \label{fig:logoRename}
\end{wrapfigure}

A label on the canvas may be clicked to enact a \toolName{Rename} refactoring on the variable (\autoref{fig:logoRename}). After renaming the \verb+squareLineLine+ variable to \verb+logo+, she then, with the list still selected, invokes the \toolName{Abstract} tool to construct a function that returns the selected item, namely the list \verb+[square1, line1, line2]+. The \toolName{Abstract} tool performs an ``Extract Method'' refactoring. Definitions (with free variables) used only in the construction of \verb+[square1, line1, line2]+ are heuristically pulled into the new function, producing a function named \verb+logoFunc+ parameterized over the remaining free variables \verb+y+, \verb+x+, \verb+w+, \verb+color+, and \verb+strokeWidth+:

\begin{lstlisting}[numbers=none, xrightmargin=.0\textwidth]
...
logoFunc y x w color strokeWidth =
  let topLeft = [x, y] in
  let square1 = square 140 topLeft w in
  let y2 = y + w in
  let line1 = line color strokeWidth topLeft [ x+ w, y2] in
  let line2 = line color strokeWidth [x, y2] \
                [ (2! * x + w)/ 2!, (2! * y + w) / 2!] in
  [square1, line1, line2]

logo = logoFunc y x w color strokeWidth
...
\end{lstlisting}

\begin{wrapfigure}{r}{0pt}
  \includegraphics[width=1.5in]{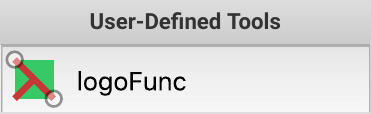}
  \caption{Custom drawing tool via type inference.}
  \label{fig:logoUserDefinedTool}
\end{wrapfigure}

The programmer has created a reusable function, which she could manually call multiple times in the program. Conveniently, however, type inference in \sns{} notices that this function accepts an $(x,y)$-coordinate and a width, so this function appears in the tool box on the right-hand side as a custom drawing tool (\autoref{fig:logoUserDefinedTool}).\footnote{Type inference mechanisms to distinguish between $x$ and $y$ coordinates, distances, and other numeric values are described in the appendix included in the \suppMaterials{}.} User-defined drawing tools behave the same way as the standard library drawing tools such as the square and line tools already demonstrated: drawing on the canvas with the tool will insert a new call to the custom function into the program.

\subsection{Refactoring}
\label{sec:refactoring}

Although the programmer has produced a reusable design, she may change her mind about particular details in the program. For example, the programmer may like to clean up the code and also may want to add a border to the design. \sns{} provides features to help implement code changes such as these two.

Similar to list widgets, \emph{call widgets}---drawn with solid borders---are placed around items that are produced by function calls in the program. The programmer clicks the call widget for the first call to \verb+logoFunc+, which (a) exposes the function's arguments above the called widget, and (b) focuses any changes to just that function. The rest of the program output (there is none currently) disappears from the canvas for the duration of the focused editing session.

\begin{wrapfigure}{r}{0pt}
	\includegraphics[width=2in]{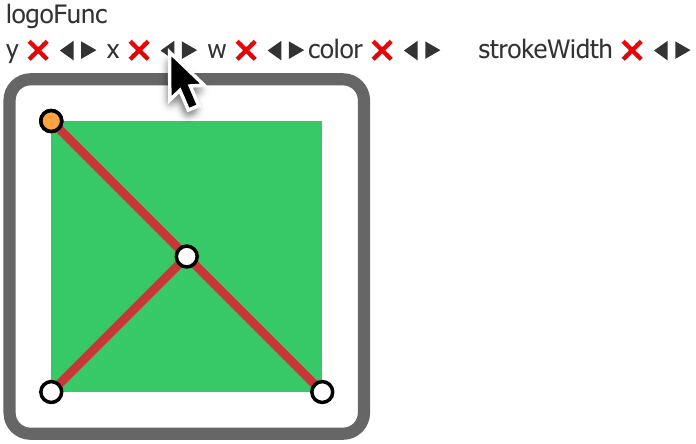}
  \caption{Focused editing.}
  \label{fig:logoFuncFocused}
\end{wrapfigure}

When a function is focused, its arguments may be renamed, removed, or reordered; an argument may be added by selecting a shape feature within the function and choosing \toolName{Add Argument} from the output tools menu. In this case, the programmer clicks the \toolName{Reorder Argument} arrows to place the \verb+x+ argument before \verb+y+.

Now the programmer would like to add a border. After changing the two existing lines to black via color sliders, she then snap-draws a polygon to the four corners of the square with the ``Polygon'' tool from the toolbox. Because she is focused on this function, the new polygon, called \verb+polygon1+, is added to the return expression of the \verb+logoFunc+ function instead of to the program's final shape list (see lines 20-25 of \autoref{fig:logoFinalCode}). With the polygon selected, she uses the polygon's sliders to set its fill color to transparent and to adjust its stroke width.

The lambda logo program is now completed to the programmer's satisfaction; the final code is shown in \autoref{fig:logoFinalCode}. The programmer was able to create the program entirely through output-directed manipulations on the canvas by using tools for \emph{drawing}, \emph{relating}, \emph{abstracting}, and \emph{refactoring}.

\begin{figure}[h]
  \begin{lstlisting}[numbers=left,numbersep=7pt,xleftmargin=.01\textwidth, xrightmargin=.0\textwidth]

y = 127

x = 158

w = 156

color = 362

strokeWidth = 5

logoFunc x y w color strokeWidth =
  let topLeft = [x, y] in
  let square1 = square 140 topLeft w in
  let y2 = y + w in
  let xYPair = [ x+ w, y2] in
  let line1 = line color strokeWidth topLeft xYPair in
  let line2 = line color strokeWidth [x, y2] \
                [ (2! * x + w)/ 2!, (2! * y + w) / 2!] in
  let polygon1 =
    let pts = [[x, y], [ x+ w, y], xYPair, [x, y2]] in
    let [color, strokeColor, strokeWidth] = \
          [529, 360, 5] in
      polygon color strokeColor strokeWidth pts in
  [square1, line1, line2, polygon1]

logo = logoFunc x y w color strokeWidth

svg (concat [
  logo
])
\end{lstlisting}

  \caption{The final code for the lambda logo example (\autoref{fig:logoFinal}), \mbox{produced} entirely by output-directed manipulations on the canvas.}
  \label{fig:logoFinalCode}
\end{figure}

\section{Design and Implementation}
\label{sec:design-and-implementation}

\sns{} is a browser application written in the functional language Elm~\cite{Elm} extended with custom providers for mutation and exceptions.

\begin{figure}
\centering
\includegraphics[width=3.00in]{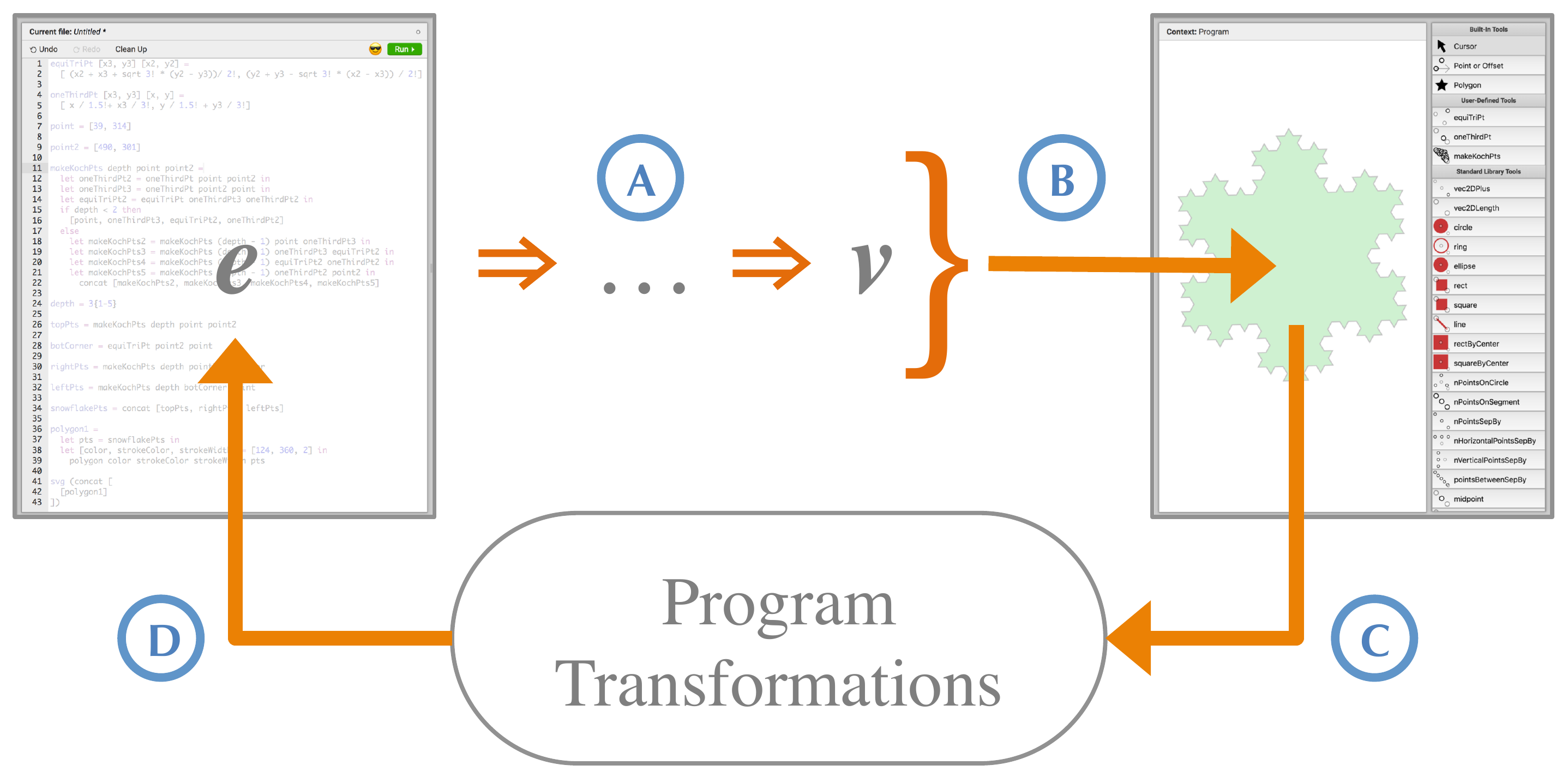}
\caption{Output-directed programming workflow. %\\[4pt]
(A) Program $e$ evaluates to value $v$ of type SVG. %\\[4pt]
(B) Graphical widgets for manipulating
the output value, intermediate values, and the program. %\\[4pt]
(C) Programmer selects from toolbox (\cf{}~\autorefFigTwoPaneUI{}a) and draws on
canvas, or manipulates existing shapes and selects a tool from the floating
tools menu (\autoref{fig:outputToolsPanel}). %\\[4pt]
(D) One or more candidate programs are produced; the programmer previews
the changes and then chooses one
(\autoref{fig:outputToolsPanel}).}

\label{fig:workflow}

\end{figure}

\autoref{fig:workflow} depicts the workflow for output-directed programming
in \sns{}.
The programmer-facing language for users is a standard functional
language with Elm-like syntax.
A text-based program written in this language is evaluated, and its SVG output value is
rendered on the output canvas. The output is overlaid with widgets for selecting and manipulating sub-values (discussed below). The programmer may draw a shape, or drag an item to change a number in the program~\cite{sns-pldi}, or make a selection and choose a program transformation from the floating tools menu (\autoref{fig:outputToolsPanel}). The resulting modified program is placed in the code box and is re-evaluated producing new output on the canvas.

Most of \sns{}'s tools (summarized in \autoref{fig:toolsTable}) operate on selections on the canvas. Different kinds of sub-values of the final output may be selected: whole shapes may be selected, or sub-features of shapes may be selected---namely, positional properties (via points rendered on the corners and edge midpoints of shapes), size properties (via selection regions for width and height), colors (via color sliders), or strokes (via stroke width sliders).
When the programmer makes a selection and then invokes a transformation, the selected sub-values are ``arguments''
to the transformation and are interpreted in a transformation-specific way.

\newcommand{\newTool}
  {\blackStar}
\newcommand{\oldTool}
  {\textcolor{white}{\blackStar}}
  %% TODO not sure why phantom screws up layout...
  %% {\phantom{\blackStar}}

\newcommand{\toolInfo}[6]
  {#2 \toolName{#1} & #3}

\newcommand{\uiInfo}[6]
  {#2 \toolName{#1} (UI) & #3}
  %% {#2 \toolName{#1} (Mostly UI Concern)}

\begin{figure}[t]
\small
\centering
\setlength\tabcolsep{0.015in}
\begin{tabular}{p{0.387\columnwidth}r}
\hline
%% \textbf{Draw} & \#Ex \\ \hline
\textbf{Drawing} & \#Ex \\ \hline
\toolInfo{Draw Shape}{\oldTool}{16}{}{}{} \\
%% \toolInfo{Draw Custom Shape (``\toolName{Lambda}''~\cite{sns-uist})}{\oldTool}{}{}{}{} \\ 
\toolInfo{Draw Custom Shape}{\oldTool}{4}{}{}{} \\ 
\toolInfo{Dupe}{\oldTool}{2}{}{}{} \\
\toolInfo{Draw Point}{\newTool}{3}{}{}{} \\
\toolInfo{Draw Offset}{\newTool}{9}{}{}{} \\
\toolInfo{Delete}{\newTool}{4}{}{}{} \\
\toolInfo{Focused Drawing}{\newTool}{1}{}{}{} \\
\uiInfo{Snap Drawing}{\newTool}{15}{}{}{} \\
\\ \hline
%
%% \textbf{Relate} & \#Ex \\ \hline
\textbf{Relating} & \#Ex \\ \hline
\toolInfo{Make Equal}{\oldTool}{12}{}{}{} \\
\toolInfo{Relate}{\newTool}{2}{}{}{} \\
\uiInfo{Distance Features}{\newTool}{6}{}{}{} \\
\\
\\
\\
\end{tabular}
\hfill
\begin{tabular}{p{0.459\columnwidth}r}
\hline
%% \textbf{Group/Abstract/Repeat} & \#Ex\\ \hline
\textbf{Abstracting} & \#Ex\\ \hline
\toolInfo{Group}{\oldTool}{8}{}{}{} \\
\toolInfo{Abstract}{\oldTool}{9}{}{}{} \\
\toolInfo{Merge}{\oldTool}{0}{}{}{} \\
\toolInfo{Repeat over Function Call}{\newTool}{2}{}{}{} \\
\toolInfo{Repeat over Existing List}{\newTool}{3}{}{}{} \\
\toolInfo{Repeat by Indexed Merge}{\newTool}{2}{}{}{} \\
\toolInfo{Fill Hole}{\newTool}{2}{}{}{} \\
\uiInfo{List Widgets}{\newTool}{9}{}{}{} \\
\\ \hline
%
%% \textbf{Refactor} & \#Ex \\ \hline
\textbf{Refactoring} & \#Ex \\ \hline
\toolInfo{Focus Expression}{\newTool}{3}{}{}{} \\
\toolInfo{Rename in Output}{\newTool}{15}{}{}{} \\
\toolInfo{Add/Remove/Reorder Arg.}{\newTool}{6}{}{}{} \\
\toolInfo{Reorder List}{\newTool}{4}{}{}{} \\
\toolInfo{Add to Output}{\newTool}{1}{}{}{} \\
\toolInfo{Select Termination Cond.}{\newTool}{1}{}{}{}
\end{tabular}
\caption{Program transformations and user interface features (UI) \mbox{in \sns{}}.
Those marked with \newTool{} are new to our system; the remainder are improvements upon \citeSnsUist{}. The \#Ex column indicates the number of examples in \autoref{fig:examples} in which the feature was used.}
\label{fig:toolsTable}
\end{figure}

Looking forward to applying ODP to larger programs, manipulation of the final output alone will be insufficient. The output by itself does not represent the process by which the output was produced. We believe some of the intermediate process should be exposed on the canvas for manipulation.

Therefore, in addition to the sub-values of the final output described above, the programmer may also select two additional kinds of terms associated with the program's execution.
First, we also expose widgets on the canvas that correspond to \emph{intermediate
values} produced during execution.
Second, we allow the programmer to select a \emph{sub-expression} in the program
to denote the context under focus.
These additional kinds of selections---intermediate values and focused
sub-expressions---serve as further arguments to program transformations
and enable the programmer to modify portions of the computation that
do not have a direct representation in the final output.

To highlight the key ideas behind our implementation, we discuss these two novel selection types below. We also discuss the degree to which several of our transforms are ordinary automatic refactorings  (i.e. not specific to SVG), as well as the generic provenance tracing that the transformations rely upon to associate canvas selections with program locations.

\subsection{Intermediate Value Widgets}

To allow manipulation of computation steps before the final output, \sns{} displays widgets for three kinds of intermediate values produced \emph{during} execution: point values, offset values, and list values.

Points (number-number pairs, \eg{}~\verb+[10, 20]+) produced at intermediate execution steps are exposed for manipulation as \emph{point widgets} on the canvas. Between selected points, we also draw \emph{distance features} for selecting Cartesian distances.

In graphics code, it is common to define offsets from some base $x$ or $y$ values. Therefore, during execution, when a numeric amount is added to or subtracted from an $x$ or $y$ coordinate, an \emph{offset widget} is drawn on the canvas as a horizontal or vertical arrow from the initial point, where the length of the arrow is the numeric amount of the offset. The arrow itself may be selected or dragged to change the offset amount.

\begin{figure}[bt]
  \begin{wrapfigure}{r}{0pt}
    \includegraphics[width=1.2in]{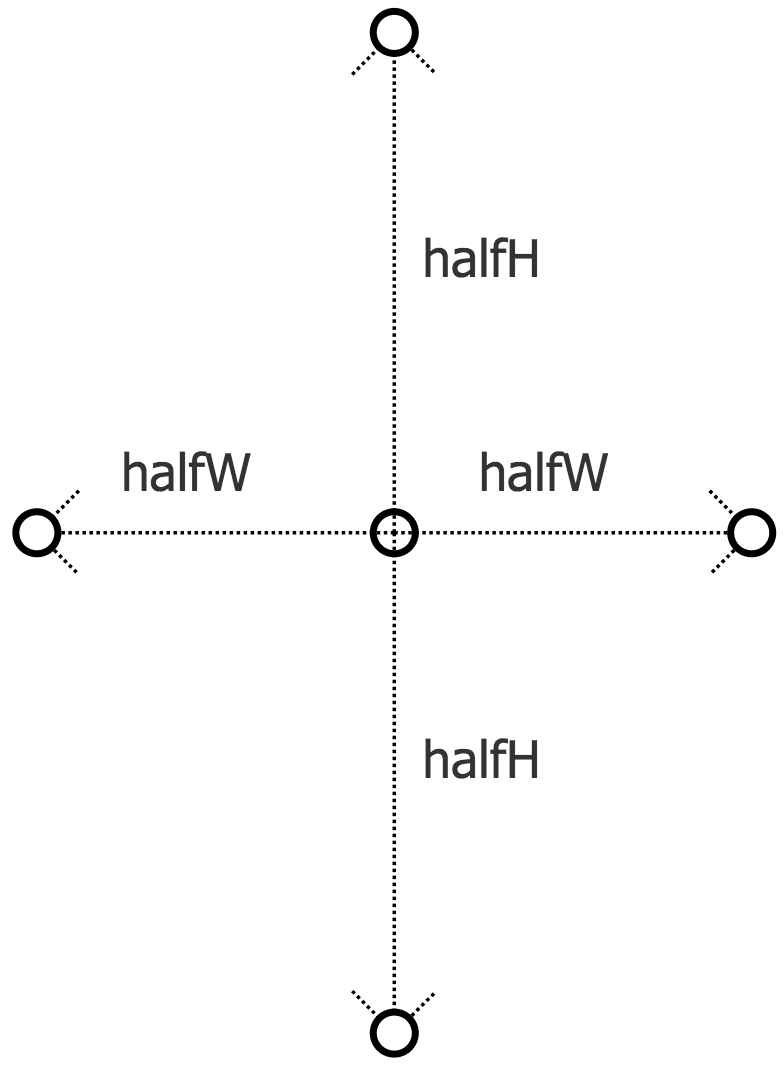}
  \end{wrapfigure}
\begin{lstlisting}
[x, y] as point = [208, 256]

halfW = 102  

xOffset = x + halfW

xOffset2 = x - halfW

halfH = 145 

yOffset = y - halfH

yOffset2 = y + halfH
\end{lstlisting}
  \caption{Point and offset widgets are displayed based on program execution to allow manipulation of intermediate computations. Depicted is an early step in the creation of \autoref{fig:examples}xii.}
  \label{fig:rhombusSkeleton}
\end{figure}

To complement the new point widgets and offset widgets, a ``Point and Offset'' tool in the toolbox (\autorefFigTwoPaneUI{}b) allows the programmer either to click on the canvas to add a point definition to the program (\eg{} \verb+[x, y] as point = [10, 20]+) or to drag on the canvas to add a new offset definition (\eg{} \verb-xOffset = x + 5-). Offsets not drawn from an existing point also insert a new point for the base of the offset. To ease the creation of symmetry, offset amounts may snap to each other while drawing. The symmetry is enforced by introducing a shared variable for offset distance.
For example, the program in \autoref{fig:rhombusSkeleton} is rendered on the canvas as a point widget from which four offsets emanate. The variables \verb+halfW+ and \verb+halfH+ are shared offset amounts produced by snap-drawing (and subsequently renamed by the programmer). The ends of the offsets can serve as snap targets for future drawing (\eg{} the rhombuses in \autoref{fig:examples}xii). Note that \autoref{fig:rhombusSkeleton} was created by the using the ``Point and Offset'' tool, but the rendered widgets would be the same if the program were instead written by text edits---only the code matters. \sns{} emits widgets from the \emph{evaluator} when appropriate computation patterns are encountered.

A third type of intermediate widget---a \emph{list widget}---is emitted whenever an execution step evaluates to a list of graphical items. A list widget is drawn as a dashed box encompassing the items (\autoref{fig:logoListWidgetSelected}). Any list of shapes, list of points, or list of lists in the program can thus be selected. List widgets obviate the need for a graphics-specific construct to denote groups of shapes---a group is just a list of shapes, and the list widget facilitates selection of, and thus operations on, the group.

\subsection{Expression Focusing}

To enable manipulation of subsections of the program, a program expression may be selected to control the syntactic scope of
changes made to the code. Most notably, focusing a shape group (\ie{} list of shapes) or a function causes drawing operations to insert shapes into that group or function rather than the final output. Accordingly, a recursive design can be created by drawing a function inside itself.

When a function call produces graphical elements, a \emph{call widget} is displayed around the returned elements as a box with a gray solid border. Clicking the gray border \emph{focuses} the function call (\autoref{fig:logoFuncFocused}). The remainder of the program's output is hidden and drawing operations are interpreted inside the focused function. Clicking the call widget again or pressing the ``escape'' key returns focus back to the entire program.

Literal (non-function) definitions may also be focused. If a canvas selection refers to the right hand side of an assignment, a \toolName{Focus Expression} tool is offered to focus that assignment.

\subsection{Generic Refactoring}

Looking forward to applying ODP to domains beyond vector graphics, a number of \sns{}'s tools apply standard automatic refactorings.

Most notably, labels are drawn next to most widgets on the canvas to aid comprehension---any label may be clicked to \toolName{Rename} the associated variable and its uses.

\sns{} also includes tools for refactoring functions. The \toolName{Abstract} tool performs a generic ``Extract Method'' refactoring, building a new function parameterized over the expression's free variables. Also, when a function call is focused by clicking its call widget, \sns{} offers actions to add, remove, or reorder arguments. Finally, the \toolName{Merge} tool performs clone elimination.

\subsection{Provenance Tracing}

While the widgets for intermediate values expose relevant steps of the computation on the canvas, occasionally the transformations require knowledge about parts of the computation that are not directly represented by widgets. For example, the \toolName{Add Argument} tool searches for every expression that affected the selected value and separately offers each such expression as a possible new argument to the function.

To help the tools answer these kinds of questions, the \sns{} evaluator performs tracing on \emph{every} execution step: the resultant value is tagged with the expression being evaluated as well as pointers to the prior (tagged) values used in the immediate computation; transitively following these prior tagged values reveals the dependencies of the computation. To answer containment queries, we additionally add pointers from list elements to the list(s) containing them. Our tracing discards certain control flow information, most notably pattern match de-structuring operations; future versions of \sns{} may adopt the Transparent ML~\cite{aCoreCalculusForProvenance} tracing scheme to preserve this information. Neither our tracing nor Transparent ML is specific to SVG---we suspect generic tracing will be useful for ODP in multiple domains.

\sns{} becomes sluggish on larger examples.
Although tracing theoretically adds memory and time overhead to evaluation, the culprit is not tracing itself but is usually comparison between large traces or widget processing.

\section{Case Study of ODP Examples}
\label{sec:examples}

\newcommand{\maybeEmph}[1]
  {\emph{#1}}
  %% {#1}
\newcommand{\maybeBold}[1]
  {\uppercase{\textsl{#1}}}
  %% {#1}
%
\begin{figure*}[bt]
  \includegraphics[width=\textwidth]{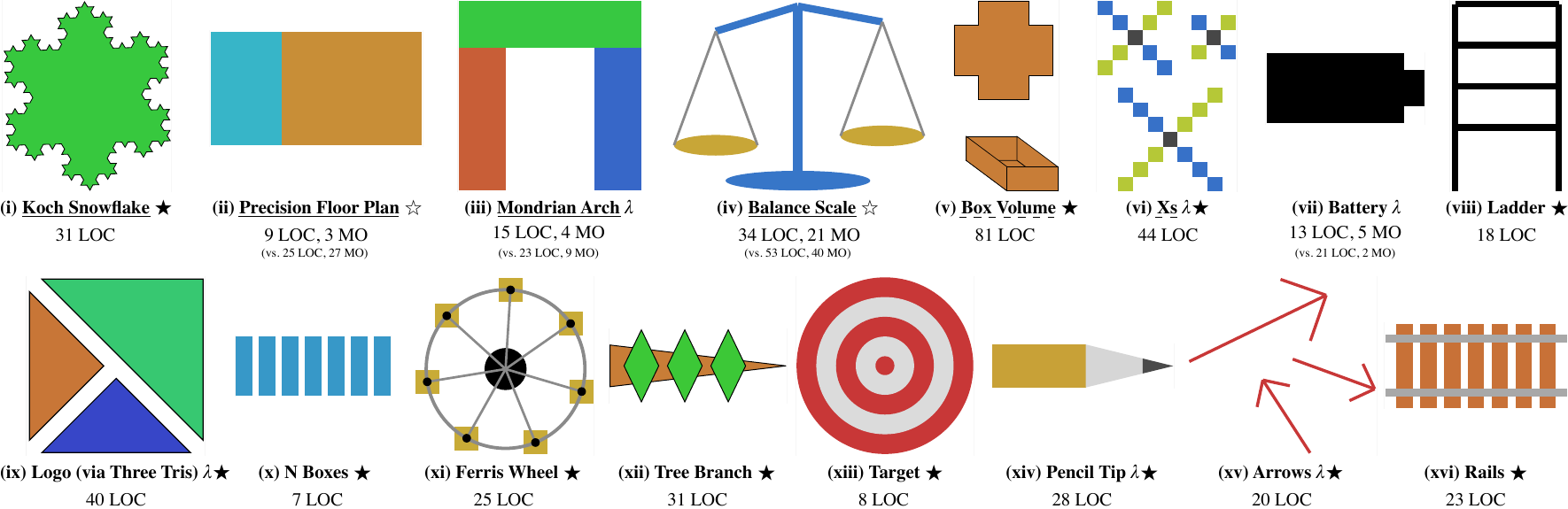}
  \caption{
Example programs created solely via output-directed manipulations. %%\\[4pt] %% NOT ALLOWED!
LOC = Source lines of code.
$\lambda$ = Final design is a function that appears as a drawing tool. %%\\[4pt]
\maybeBold{Comparison to \citeSnsUist{}:}
If an example can similarly be created in \citeSnsUist{}, code metrics for \citeSnsUist{} appear in parentheses. Math operations (MO) indicate code simplifications due to our improvements.
\blackStar{} (resp. \openStar{}) =
Task cannot (resp. can with undesired parameterization) be created
\mbox{in \citeSnsUist{}}. %%\\[4pt]
\maybeBold{Source of examples:} %%\\
\wwidPbd{}~\citeWWIDBenchmarks{}: Tasks marked with \underline{underline}
(\underline{d}a\underline{s}h\underline{e}d = only partially completed). %%\\
\mbox{Lillicon \citeLillicon{}:} \mbox{\maybeEmph{(vii) Battery}}.
\mbox{QuickDraw \citeQuickDraw{}:} \maybeEmph{(viii) Ladder}.
\mbox{\sns{} \citeSnsUist{}:} \maybeEmph{(ix) Logo (via Three Tris)}.
\mbox{\sns{} \citeSnsPldi{}:} \maybeEmph{(x) N Boxes}, \maybeEmph{(xi) Ferris Wheel}.
}
  \label{fig:examples}
\end{figure*}

To explore the expressiveness of ODP in \sns{}, we implemented \numExamples{} parametric designs summarized in \mbox{\autorefFigExamples{}}.
Drawn from various sources~\cite{sns-pldi,sns-uist,WWIDbenchmarks,Lillicon,QuickDraw2012}, the designs exercise different features: \mbox{6 designs} are parameterized functions that appear as drawing tools at the end of construction, 6 involve repetition by position, 1 involves repetition by other features (radius and color, \autorefFigExamples{}xiii), and 1 uses recursion (\autorefFigExamples{}i).
All \numExamples{} programs, spanning \numLoc{} lines of code total, were built entirely via output-directed manipulations, without any text editing in the code box.

Below, we discuss common patterns in how we used the tools to implement the examples. Next, we recount the key steps in building the recursive Koch curve and explain the process to construct designs involving repetition.
Finally, we describe limitations by discussing
the tasks from the \wwid{}~\cite{WWIDbenchmarks} benchmark suite that the current system cannot complete without text edits.

\subsection{Authoring}

The lambda logo walkthrough in the overview framed the \sns{} workflow as a four stage \emph{draw-relate-abstract-refactor} authoring process. In practice, design construction cannot always be cleanly delineated into precise stages in a fixed order. For example, while constructing the designs of \autorefFigExamples{}, we often specified the relationships \emph{before} drawing the shapes: we laid out the desired parameterization of the design using points and offsets (as in, \eg{}, \autoref{fig:rhombusSkeleton}) and afterwards attached shapes using snap-drawing.

The \#Ex column of \autoref{fig:toolsTable} lists how many of the \autorefFigExamples{} examples utilized each tool. Indeed, besides shape drawing and variable renaming, the most widely used functionality was \sns{}'s snap-drawing ability---not surprising given that the goal was to create parametric designs.

It is notable that to encode spatial constraints we more often preferred to snap-draw rather than use the \toolName{Make Equal} tool. \toolName{Make Equal} is more flexible, but not only does it require extra clicks compared to snap-drawing, \toolName{Make Equal} can offer a large number of different but hard-to-distinguish ways to enforce a constraint (\autoref{fig:outputToolsPanel} is a tame example). To help, \sns{} employs a ranking heuristic that seems to work well in practice: \toolName{Make Equal} prefers changes that rewrite terms near each other and later in the program. The least used tool for specifying relationships was the \toolName{Relate} tool. \toolName{Relate} guesses a mathematic relationship between selected items---we only used it for constraints involving thirds.

Offsets plus snaps and \toolName{Make Equal} was sufficient, but not always convenient, for creating reasonable parameterizations of the designs. Laying out offsets beforehand requires forethought. A future \sns{} might benefit from tools to break constraints or invert dependencies after the fact.

As indicated in \autoref{fig:toolsTable}, we did not use the \toolName{Merge} tool and three tools were used only once, all on the most challenging example, the Koch snowflake. The \toolName{Merge} tool merges multiple copies of a shape into a function---we prefer \toolName{Abstract} instead because it requires only a single example. The three tools only used for the Koch fractal all played a role in the workflow to specify recursion, which we now recount.

\subsection{Recursive Koch Snowflake}

We highlight the key steps in creating the recursive von Koch snowflake design~\cite{kochFractal} (\autorefFigTwoPaneUI{}, \autorefFigExamples{}i). This task requires manipulating intermediates---program has no output throughout most of its development---and focused editing is needed to build the recursive function. Here we emphasize the steps to specify the recursion, although the video in the \suppMaterials{} walks through the entire construction.

Each side of the Koch snowflake is based on a recursive motif shown in \autorefFigTwoPaneUI{}f. The motif requires two helper functions: \mbox{a function} that, given two points, computes a point $\frac{1}{3}$ of the way between them (\verb+oneThirdPt+, \autorefFigTwoPaneUI{}d), and a function that, given two points, computes a third point that completes an equilateral triangle (\verb+equiTriPt+, \autorefFigTwoPaneUI{}e). The two helpers are created without text edits via the \toolName{Relate} and \toolName{Make Equal} tools, respectively. Each helper takes in two points and is thus exposed as a drawing tool (\autorefFigTwoPaneUI{}c).

\begin{wrapfigure}{r}{1.6in}
    \centerline{\includegraphics[width=1.4in]{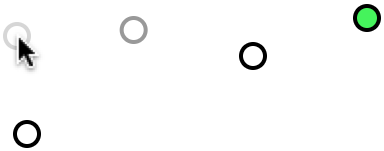} \vspace{0.5em}}
	\includegraphics[width=1.6in]{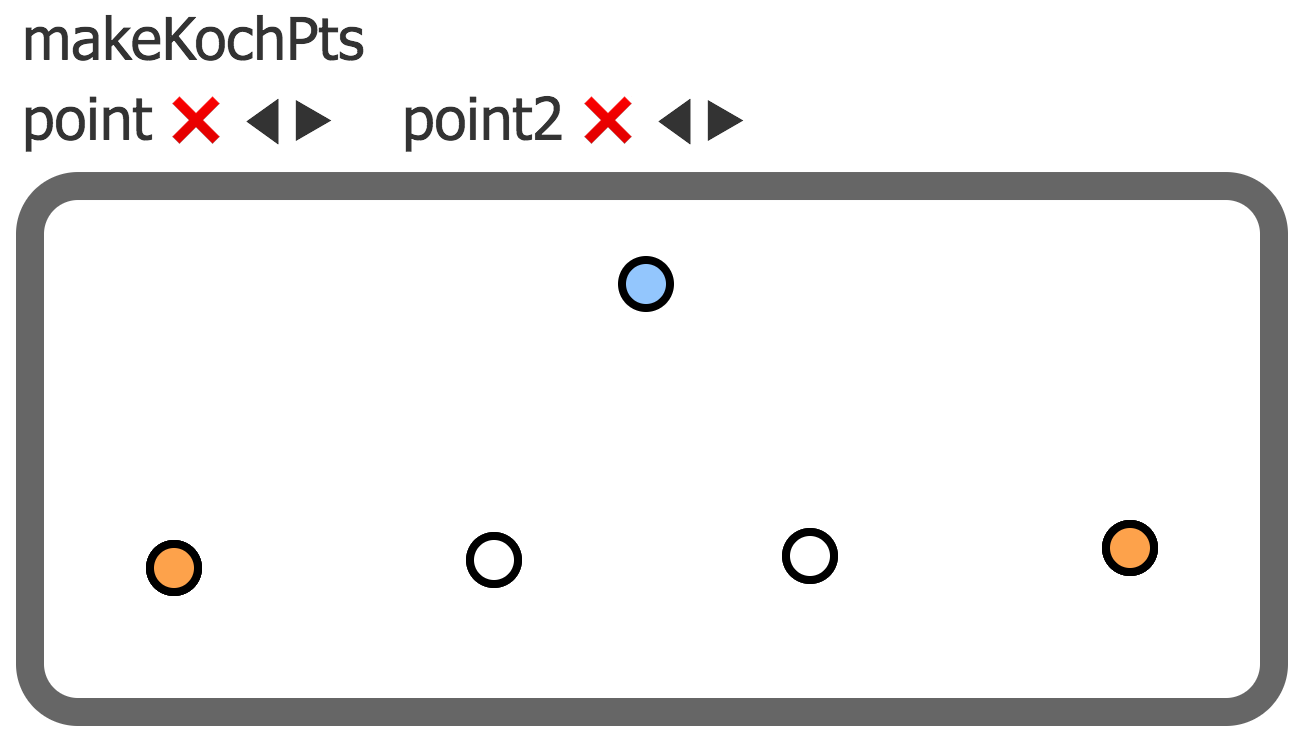}
	\caption{Initial Koch motif.}
    \label{fig:initialKochMotif}
\end{wrapfigure}

The Koch motif is created by snap-drawing \verb+oneThirdPt+ forward and backwards to yield the $\frac{1}{3}$ and $\frac{2}{3}$ points; \verb+equiTriPt+ is used to create a point equidistant from the $\frac{1}{3}$ and $\frac{2}{3}$ points. The bottom of \autoref{fig:initialKochMotif} shows this initial motif after abstraction. Note that when a function is focused, as in \autoref{fig:initialKochMotif}, input points are rendered in orange and output points in blue. The code is not recursive yet:

\begin{lstlisting}
makeKochPts point point2 =
  let oneThirdPt2 = oneThirdPt point point2 in
  let oneThirdPt3 = oneThirdPt point2 point in
  equiTriPt oneThirdPt3 oneThirdPt2 
\end{lstlisting}

Drawing the function inside itself causes \sns{} to insert a recursive skeleton and the recursive call:

\begin{lstlisting}
makeKochPts point point2 =
  let oneThirdPt2 = oneThirdPt point point2 in
  let oneThirdPt3 = oneThirdPt point2 point in
  let equiTriPt2 = equiTriPt oneThirdPt3 oneThirdPt2 in
  if ??terminationCondition then
    equiTriPt2
  else
    let makeKochPts2 = makeKochPts point oneThirdPt3 in
      equiTriPt2
\end{lstlisting}

To avoid infinite recursion, the temporary guard expression \verb+??terminationCondition+ evaluates to \verb+false+ the first time the function is encountered, and \verb+true+ if the function has appeared earlier in the call stack, affecting termination at a fixed depth. The termination condition may be selected later.

\begin{wrapfigure}{r}{0pt}
	\includegraphics[scale=0.2]{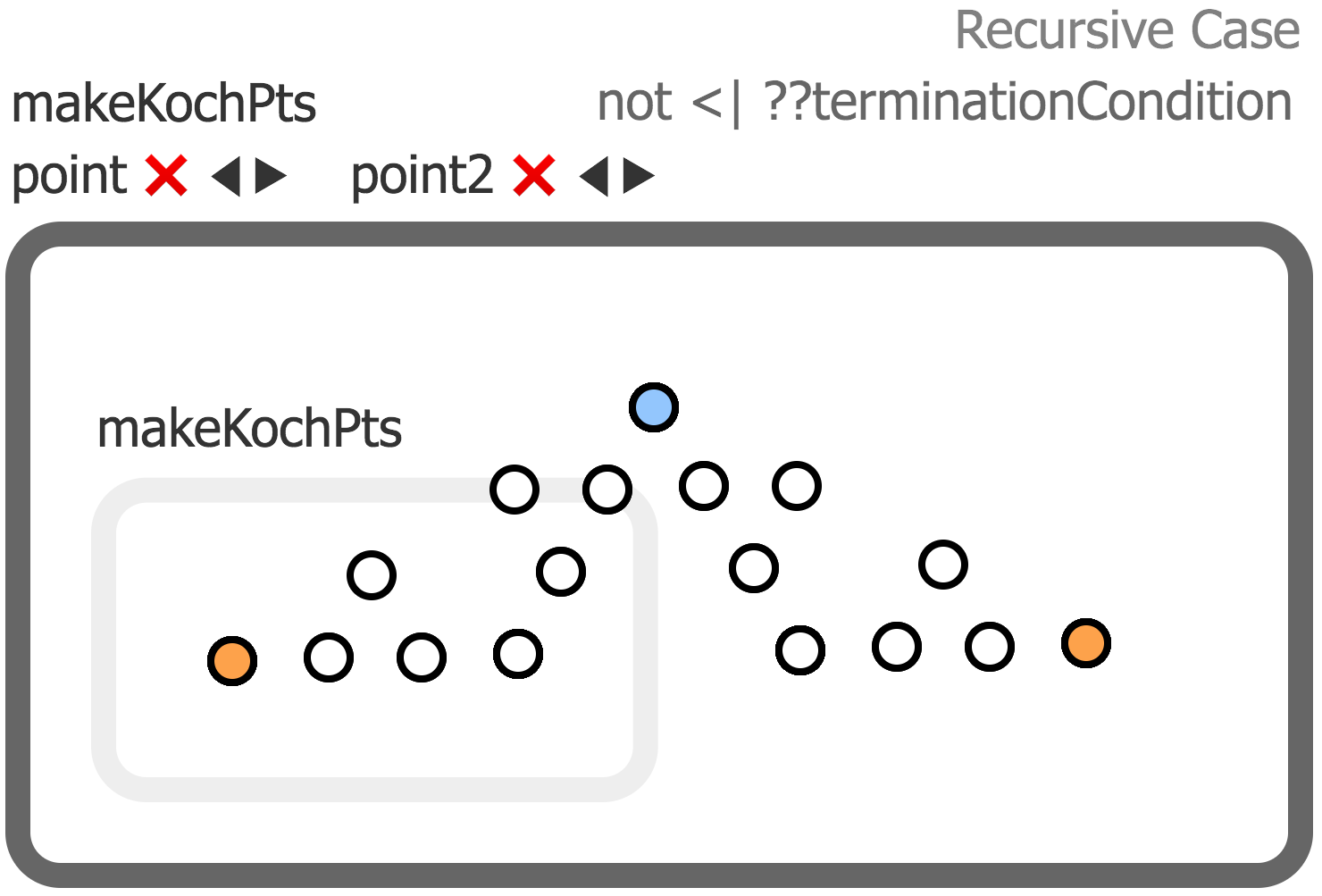}
	\caption{Recursive case.}
	\label{fig:kochRecursiveCase}
\end{wrapfigure}

\begin{wrapfigure}{r}{0pt}
	\includegraphics[width=1.15in]{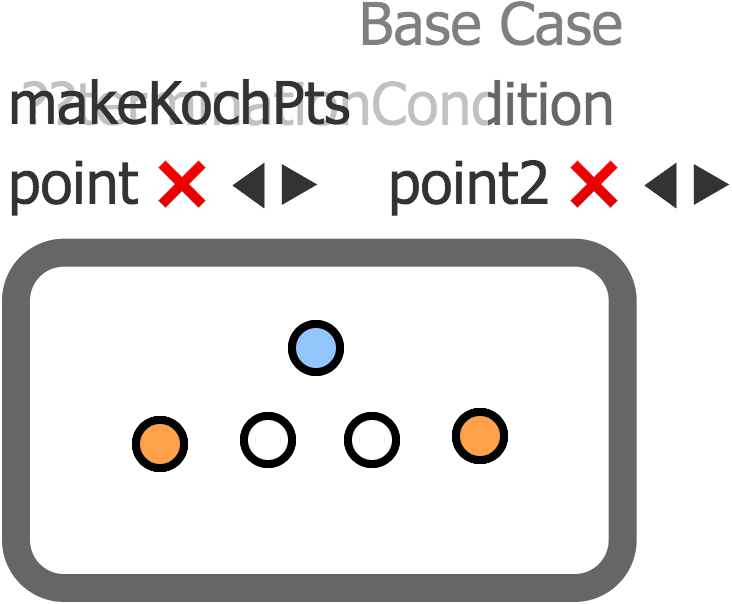}
	\caption{Base case.}
	\label{fig:kochBaseCase}
\end{wrapfigure}

After drawing the function between the remaining three pairs of points the design looks like a fractal (\autoref{fig:kochRecursiveCase}). However, only one point is output from the function---the white points are only intermediates. Moreover, most of those intermediates are inside calls to the base case.
Although focused on the function's recursive case, to modify the output of the base case, the programmer can click the call widget of a call to the base case (the inner light gray border in \autoref{fig:kochRecursiveCase}) to focus that call instead (\autoref{fig:kochBaseCase}).

The \toolName{Add to Output} and \toolName{Reorder in List} tools are used to place the selected points into the output of the base case (resulting in the list on line 16 of \autorefFigTwoPaneUI{}). Once complete, the recursive case may be refocused, and, using the same two tools, its output is specified to be the concatenation of the lists returned from calls the the base case (line 22 of \autorefFigTwoPaneUI{}).

What remains is to choose a termination condition. The focused call widget for \verb+makeKochPts+ displays the conditional \verb+not <| ??terminationCondition+ for the recursive case. Clicking the conditional offers termination conditions. Currently, \sns{} implements only one: fixed depth, in which a \verb+depth+ argument is decremented on the recursive calls. These changes are visible in the final code (\autorefFigTwoPaneUI{}f).

With all the points for the Koch snowflake now on the canvas, a polygon is attached to them by selecting the ``Polygon'' tool and clicking the list widget to select the points. With widgets hidden, the final snowflake is shown in \autorefFigTwoPaneUI{}.

\begin{figure}[t]
\begin{lstlisting}
rhombusFunc2 ([x, y] as point) =
  let halfW = 40 in
  let halfH = 83 in
  rhombusFunc point halfW halfH

repeatedRhombusFunc2 =
  map rhombusFunc2 leafAttachmentPts 
\end{lstlisting}
  \caption{Code fragment from \autorefFigExamples{}xii produced by the \toolName{Repeat over Existing List} tool. The tool creates a function (\texttt{rhombusFunc2}) that produces a shape given a single point, and maps that function over a list of points defined elsewhere in the program (\texttt{leafAttachmentPts}).}
  \label{fig:repeatOverExistingList}
\end{figure}

\subsection{Repetition}

Of the \autorefFigExamples{} examples, 7 utilize \sns{}'s tools for creating repetitive designs.
A selected shape may be repeated either over an existing list of points in the program or over a new call to any function that returns a list of points (the default toolbox in \autorefFigTwoPaneUI{}a contains several such functions). \autoref{fig:repeatOverExistingList} illustrates code produced by these repetition tools.

For creating designs that vary over a non-spatial attribute, \sns{} alternatively offers a programming by demonstration workflow: after manually laying out the first few shapes of a repetitive design, the \toolName{Repeat by Indexed Merged} tool syntactically merges the shape expressions into a single function that is called with an argument \verb+i+ that takes on the values \verb+0+, \verb+1+, \verb+2+, \etc{} Using a simple form of sketch-based synthesis~\cite{SketchingThesis}, a special tool called \toolName{Fill Hole} resolves syntactic differences between the original shape expressions by synthesizing expressions that refer to the index variable \verb+i+. \autorefFigExamples{}x and xiii were constructed with this workflow.

\subsection{Limitations: Remaining Tasks}
\label{sec:examples-limitations}

Of the 15 tasks in the \wwid{}~\cite{WWIDbenchmarks} benchmark suite that may be interpreted as parametric drawings, our system is able to fully complete 4. What is needed to complete all the tasks without text edits?

Two of these remaining \wwidPbd{} tasks can be partially completed in this work (\autorefFigExamples{}v, vi). To fully complete them, ``Box Volume'' would require an interaction to compute and display the numeric volume of the folded box, and ``Xs'' would require more precise control over what definitions are abstracted. (Not all uses of a \verb+squareWidth+ parameter are pulled into the abstraction, causing the design to render incorrectly when drawing an X with different sized squares.)

The remaining nine tasks are diverse; no single feature would help with more than two or three. A prominent missing feature is arbitrary text boxes, with other elements placed relative to the text size. Beyond this, several examples require various list operations\hide{, albeit different such operations (\eg{} sorting, pair combinations, repetition over \emph{pairs} of points, find and replace a middle element, finding a maximum)}. \hide{Beyond minor omissions (rotation handling and drawing paths with cutouts),} \sns{} would also need to reason about intersections of lines with shape edges, to offer ways to specify overlapping and containment constraints, and to solve different kinds of such constraints simultaneously. Finally, one example would require creating \verb+if-then-else+ branches outside of a recursive or hole-filling setting.\hide{and expose a boolean flag on the canvas to swap between the branches.}

\section{Related Work}
\label{sec:related}

Many related approaches provide direct manipulation tools to (a) transform programs and/or (b) build parametric designs.

\levelThree{Parametric Computer-Aided Design (CAD)}

Feature-based parametric CAD systems record user actions as a series of steps that together act as a program encoding the creation of the design.
If an element's property is changed, dependent actions in the sequence are re-run. Among CAD systems, EBP~\cite{EBPCAD} is notable for offering a programming by demonstration workflow to create loops and conditionals.

\levelThree{Drawing With Constraints}

Several visual design systems integrate constraint specification (\eg{}~\cite{objectOrientedDrawing,paraSupportingExpressiveProceduralArtCreationThroughDirectManipulation}).
Notably, Apparatus~\cite{Apparatus}, Recursive Drawing~\cite{recursiveDrawing}, and Geometer's SketchPad~\cite{GeometersSketchpad} support recursion. Like CAD, but unlike \sns{}, these systems do not represent the design in a general-purpose, text-based programming language.

\levelThree{Constraint-Oriented Programming (COP)}

Other constraint-oriented systems, following in the footsteps of SketchPad~\cite{SketchpadThesis}, explicitly view building a constrained system as a programming task~\cite{Borning1981,Juno2,babelsbergjsABrowserBasedImplementationOfAnObjectConstraintLanguage}. While offering varying degrees of visibility into the code, these systems are distinguished by running constraint solvers alongside the program. We instead rely on standard execution semantics.

\levelThree{Programming by Demonstration (PBD)}

Several PBD approaches~\cite{WWID} use shape drawing as a domain for exploring non-textual programming techniques. These systems usually rely on a visual rather than representation of the program (\eg{} \cite{Chimera, Mondrian}) or show actions step-by-step~\cite{Metamouse}. We also use shape drawing as our application domain, but we do not hide the program text.
Although not as visual as peer PBD systems, Tinker~\cite{Tinker} is notable for supporting recursion by demonstration; indeed, any Lisp expression may be created, albeit via manipulations performed on a symbolic representation not far removed from the underlying Lisp.
More recently, PBD techniques have been developed for
data visualization~\cite{VictorDrawing},
mobile applications for collaboration~\cite{liveEndUserProgramming},
web scraping~\cite{RingerOOPSLA2016,RousillonUIST2018},
API discovery~\cite{DemoMatchPLDI2017}, and hand-drawing recognition~\cite{learningToInferGraphicsProgramsFromHandDrawnImages}.

\levelThree{Output-Directed Programming}

Several recent systems augment a regular text-based programming experience with abilities to directly manipulate {output} to enact code changes. The transformations available may be ``small'' (such as changing constants~\cite{sns-pldi,carbideAlpha,sns-oopsla},  strings~\cite{WangEtAl,LivePBE,carbideAlpha,sns-oopsla}, or list literals~\cite{sns-oopsla}), but several systems enable ``larger'' program changes via output manipulation.

Transmorphic~\cite{transmorphic} re-implements the Morphic UI framework~\cite{directnessAndLivenessInTheMorphicUserInterfaceConstructionEnvironment} but with stateless views. Transmorphic retains Morphic's ability to manipulate shapes by affecting changes to a view's (text-based) code rather than changing live object state. Adding and removing shapes as well as changing a shape's primitive properties are both supported.

APX~\cite{McDirmidAPX, McDirmidCurryOn}, like \sns{}, is a two-pane (code box and output canvas) environment for creating shape-drawing programs (APX also supports dynamic visual simulations).
On the canvas, manipulating shape attributes changes appropriate numbers in the program; a few larger changes (\eg{} grouping and inserting shapes) are also supported.
However, most of APX's interactions focus on directly manipulating terms in the code box rather than on the canvas.

\levelThree{Prior \sns{}}
\label{sec:comparison-to-sns2016}

Our prior work on \sns{}~\cite{sns-uist} introduced a \emph{draw-relate-abstract} workflow, including the ability to select sub-values in the output and invoke certain program transformations. \autoref{fig:toolsTable} indicates which tools also appeared in this prior work. In the present work, besides exposing intermediates for manipulation, offering expression focusing, and adding repetition and generic refactoring tools, we also sought to improve the prior \emph{draw-relate-abstract} tools.

The prior \sns{}~\cite{sns-uist} relied on syntactic
restrictions: shapes followed a strict ``{left-top-right-bot}'' bounding box
parameterization, special functions were needed to compose shapes, and the
program needed to be a series of top-level definitions followed by a list
literal ``main'' expression referring to the top-level definitions.
If, through text-editing, the programmer veered outside of this syntactic subset
of the language, many output-directed tools became unavailable.

In this work, we relaxed these syntactic restrictions and moved the operation of \sns{}'s tools closer to ordinary programming concepts. The drawing tools now operate based on types rather than a syntactic bounding box construct. No longer are special function calls needed to compose shapes: groups are now ordinary lists. Our generic provenance tracing enables the tools to transform shapes not defined at the top level. We also now offer multiple results for transformations.

We additionally improved code generation. Four of the examples in \autorefFigExamples{} (ii, iii, iv, vii) can also be constructed without text edits in the prior \sns{}~\cite{sns-uist}, albeit with undesirable parameterizations for two of the four (ii, iv). \autorefFigExamples{} lists the source lines of code (LOC) and math operations (MO) for the programs in both the present and prior versions of \sns{}. Reductions in LOC are incidental---our drawing tools now insert single-line function calls instead of wordier multi-line definitions. Reductions in math operations, however, do indicate better code: the prior \sns{} sometimes inserted large math expressions too complex for human comprehension. Besides discarding the bounding boxes that required properties such as shape width to be computed (\eg{} as \verb+right - left+), we also more aggressively reuse variables and simplify math in the \toolName{Make Equal} tool, aided by connecting \sns{} to an external solver (REDUCE~\cite{REDUCE}).

During our upgrades, some functionality was disabled but would be useful to restore.
Most notably, unneeded by our examples, the current \sns{} lacks a path tool. Also, tools to re-parameterize shapes \emph{en masse} could be useful, if, for example, a bounding box parameterization is desired.

\section{Conclusion and Future Work}
\label{sec:discussion}

We improved expressiveness in \sns{}, showing how to construct a variety of non-trivial
programs entirely through direct manipulation.
Our long-term goal, however, is not to create a practical drawing tool
but to use this particular application domain as a
laboratory for advancing the expressiveness of output-directed programming for a
variety of future programming settings.
There are several avenues to explore in this direction.

\levelThree{ODP for Novices}
Our work so far has assumed the \sns{} user is comfortable working in code to understand program operation. ODP interactions might also help those with little programming experience---such as domain experts or students---to quickly produce rudimentary programs. Design considerations for novices should be investigated.

\levelThree{Widget Visibility}

Because program execution may involve a large number of intermediate evaluation steps, even simple programs might clutter the canvas with widgets, making it unusable. Therefore, our current
implementation hides most widgets by default and uses heuristics to determine when to show them---generally, upon the mouse hovering over some associated shape. Additionally, widgets from intermediate expressions in standard library code---outside the visible program---are generally not shown. In the future,
these visibility choices could be more systematically controlled, \eg{}, by source code
annotations or user interface options.

\levelThree{Customizing Widgets}

Mechanisms for customization may help address the open question of how
to render graphical representations of program evaluation and of domain objects
that are not inherently visual.
One approach would be to design an API for ``\verb+toSvg+'' functions that
specify how to render graphical representations of intermediate data structures.
Analogous to ``\verb+toString+'' functions that render text, a \verb+toSvg+ API
might similarly render composite data structures via recursively calling \verb+toSvg+ for each of the structure's elements.

\levelThree{Synthesizing Program Transformations}

Each program transformation offered by \sns{} has been hand-coded one-by-one.
This effort is time-consuming, error-prone, and limits the composability of
transformations.
Program synthesis techniques have been successfully incorporated into refactoring tasks
for class-based languages~\cite{RefactoringWithSynthesis}; perhaps such
techniques can be extended to streamline the specification of output-directed transforms
that operate simultaneously on both abstract and concrete syntax.

\levelThree{User Interactions for Deciphering Human Intent}

There are often multiple valid program changes associated with a specific
user action.
To resolve ambiguity, our system displays code and output differences and asks the
user to choose.
Richer interactions are needed to explain the differences
(\eg{}, change impact analysis~\cite{Gethers:2012,Dit:2014}) and to
better resolve the user's intent (\eg{}, by asking additional
questions~\cite{Mayer2015}).

In order for the live and immediate experience afforded by
output-directed programming to scale into usable systems for
different programming tasks and user scenarios, developing and refining such
user interaction techniques remain crucial challenges to tackle in future work.

\section{Acknowledgements}

Our thanks to Philip Guo and the anonymous reviewers for helpful feedback.
This work was supported by
U.S. National Science Foundation Grant No. {1651794}.

\balance{}

%% \bibliographystyle{SIGCHI-Reference-Format}
%% \bibliography{references}

\clearpage
\section{Appendix}
\label{sec:appendix}

We
provide a few additional details about our design and implementation;
discuss examples from prior work~\cite{sns-uist}; and
describe two examples from \autoref{fig:examples} in more detail.

\newcommand{\overviewExample}[2]
  {\subsection{Example {#1}: {#2}}}

\newcommand{\toolsUsed}[1]
  {\textsf{\small Tools highlighted:} \toolName{\small #1}
   \vspace{3pt}
  }
  
\subsection{Type Inference}

Because \sns{} designs are ordinary programs, custom functions that produce shapes can be incorporated as new drawing tools in the user interface.
Any function that takes two points as input---or one point and a horizontal and/or vertical distance---is exposed as a custom drawing tool.

To determine which numbers represent horizontal/vertical distances and to provide reasonable default values for \eg{} colors and stroke widths, type inference tags each type with a set of zero or more inferred \emph{roles}. Functions in the standard library (the bottom portion of \autorefFigTwoPaneUI{}a) have role annotations in their type declarations. By using these functions, role information enters the program and propagates via type inference, usually obviating the need for manual annotation. A similar scheme called \emph{brands} (implemented via named mixin traits) is used in~\cite{McDirmidAPX}, although we additionally apply certain built-in usage rules---\eg{} that a number added to an \verb+X+ value must be a \verb+HorizontalDistance+---to infer roles in more cases.

\subsection{Value Holes}

Tagging values with provenance enables a helpful trick that we use to implement several of the new transformations (most notably the snap-drawing): to indicate that a particular program location should somehow evaluate to a particular value, we transiently insert that value as a leaf in the abstract syntax tree. Before displaying the program, each such \emph{value hole} is filled by inspecting the provenance of the value and choosing an expression that evaluates to the value (usually by using or introducing a variable). This internal workflow has been an effective way to express and resolve equality constraints within template code.

\subsection{Specifying the Focused Expression}

A focused function and call are denoted by special comments, automatically inserted into the program. Comments, unlike internal expression identifiers, are preserved across arbitrary text edits to the program.

\subsection{Completing the Examples from the Prior Sketch-n-Sketch}

Our prior work~\cite{sns-uist} presented three examples that require text edits both in the prior system and ours. Below, we discuss the additional features that would be required to complete these tasks entirely via output-directed manipulations.

Two of the remaining examples in~\cite{sns-uist} utilize paths (``Garden Logo'' and ``Coffee Mug''). Unneeded by our examples, we disabled the path tool pending internal updates to allow path drawing to work with snaps. Besides a path tool, a \toolName{Rewrite as Offsets} tool would be useful to rewrite the coordinates of selected points to be computed relative to an anchor point (akin to the domain-specific rewriting performed by the prior \toolName{Group} tool). ``Garden Logo'' would also benefit from special tooling for mirrored drawing across a vertical symmetry line. % (a mirrored design \emph{can} be created with offsets and snaps in the present work, but it's tedious).

The third remaining example (``Snip Polygon'') relied on special tooling to draw a polygon with its points positioned relative to a bounding box. That tooling was discarded along with the pervasive bounding box parameterization and would have to be restored (perhaps as a \toolName{Rewrite Relative to Bounds} tool). Additionally, to fully develop the design without text edits, \sns{} would need an output-directed mechanism for specifying \verb+min+/\verb+max+ calculations.

\newpage

\nobalance

\overviewExample{(xii)}{Tree Branch}
\label{sec:treeBranch}

\toolsUsed{Draw Offset, Repeat Over List}

To demonstrate offset widgets and repetition tools, we describe part of the tree branch task shown in \autorefFigExamples{}ii.
The overarching strategy is to build a rhombus abstracted over its center, and then repeat it over a list of points to form the leaves. We omit mentioning uses of \toolName{Rename}.

The rhombus is constructed around a central point using offsets drawn with the ``Point or Offset'' tool. Offset amounts may snap to each other while drawing, which inserts a variable for the shared offset amount (\verb+halfW+ and \verb+halfH+ below).

\begin{wrapfigure}{r}{0pt}
  \includegraphics[width=1.25in]{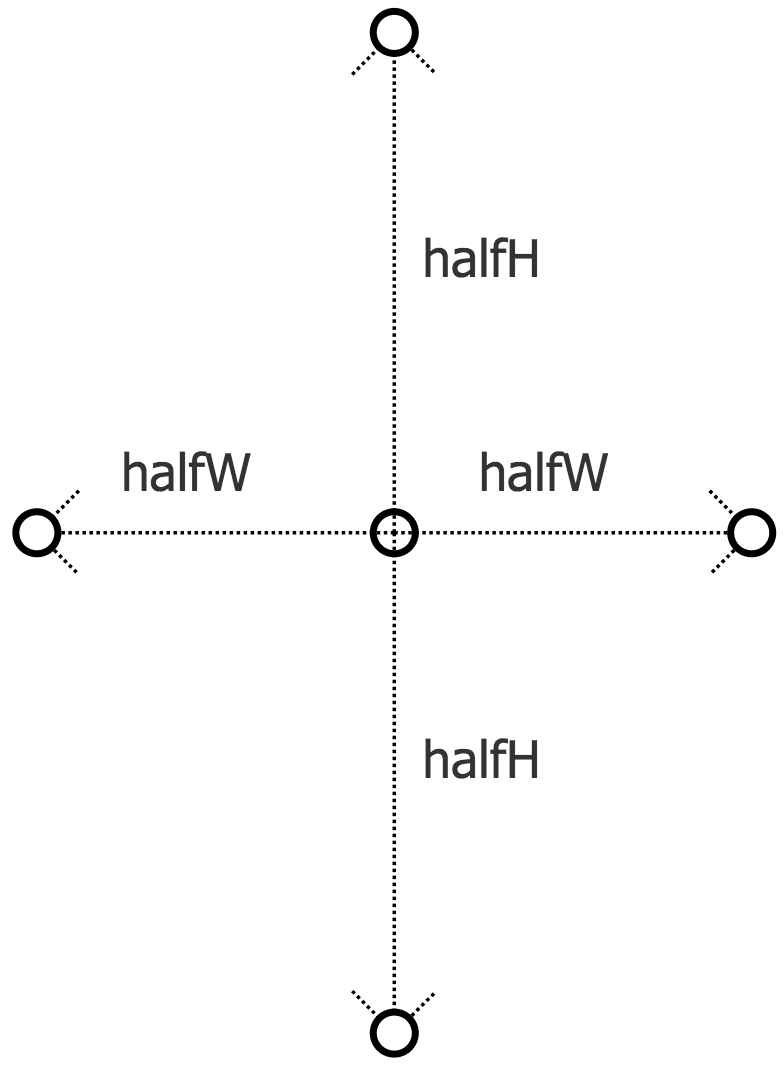}
\end{wrapfigure}

\begin{lstlisting}
[x, y] as point = [208, 256]

halfW = 102  

xOffset = x + halfW

xOffset2 = x - halfW

halfH = 145 

yOffset = y - halfH

yOffset2 = y + halfH
...
\end{lstlisting}

We then draw a polygon snapped to the offset endpoints and \toolName{Abstract} the resulting rhombus into a function parameterized over \verb+[x, y]+, \verb+halfW+, and \verb+halfH+. We will attach instances of this function over the branch.

\begin{wrapfigure}[10]{r}{0pt}
  \begin{minipage}{1.6in}
	\includegraphics[width=0.95\textwidth]{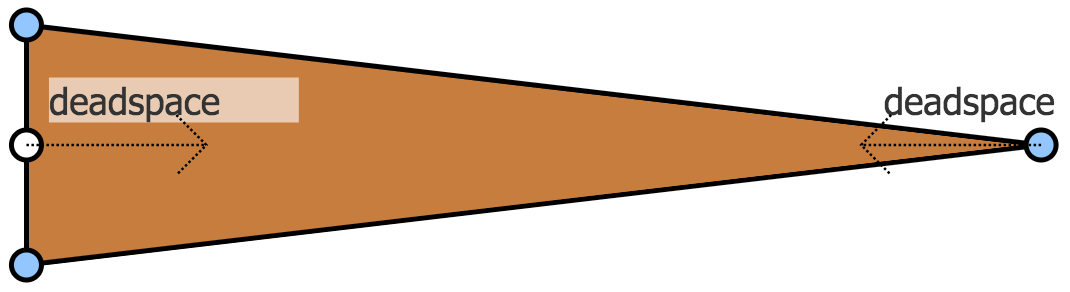}
	\includegraphics[width=0.95\textwidth]{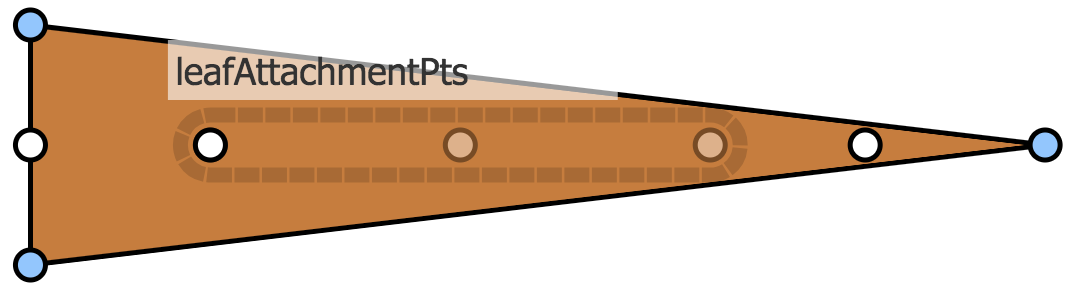}
    \includegraphics[width=0.95\textwidth]{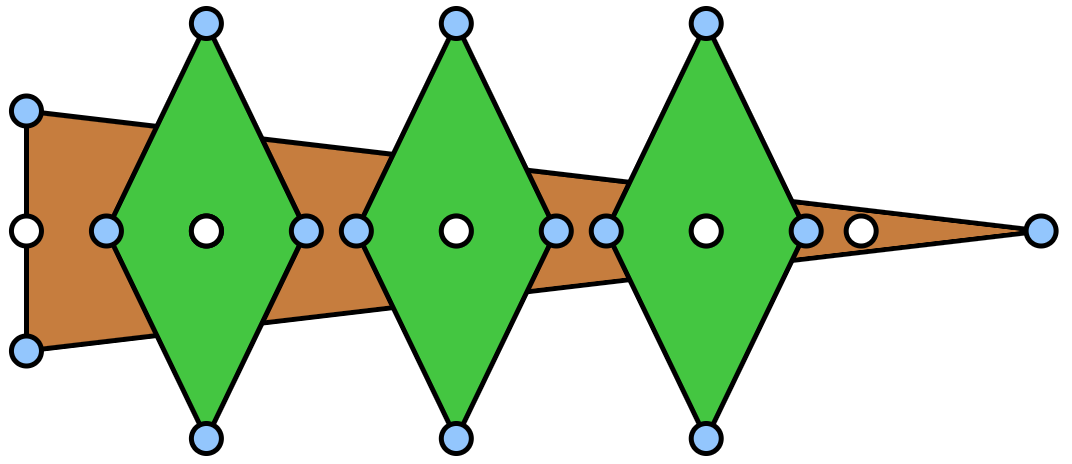}
  \end{minipage}
\end{wrapfigure}

The branch construction is shown at right. On top of a brown polygon, two ``\verb+deadspace+'' offsets are drawn inwards from the ends of the branch to form the start and end locations of the attachment points for the leaves. Between these locations we draw the \verb+pointsBetweenSepBy+ standard library function that returns points separated from their neighbors by a fixed distance. With this particular function, making our branch longer will add more leaves rather than spacing them out.

Finally, to repeat our leaf rhombus over the points, we first select the one copy of the rhombus and invoke the \toolName{Repeat Over List} tool to repeat over the attachment points we just drew. This tool creates a new function abstracted over just a single point (\verb+rhombusFunc2+ below) and maps that function over our \verb+leafAttachmentPts+, completing our leafy branch.

\begin{lstlisting}
...
rhombusFunc2 ([x, y] as point) =
  let halfW = 40 in
  let halfH = 83 in
  rhombusFunc point halfW halfH
...
repeatedRhombusFunc2 =
  map rhombusFunc2 leafAttachmentPts 
...
\end{lstlisting}

\newpage

\overviewExample{(xiii)}{Target}
\label{sec:examples-target}

\toolsUsed{Repeat by Indexed Merge, Fill Hole}

Repeating over a point list allows copies of shapes to vary in their spacial positions, but not over any other attributes such as size or color. To support other repetition scenarios where the varying attributes could be calculated from an index \mbox{(\ie{}~\texttt{0}, \texttt{1}, \texttt{2}, \dots)}, we offer a  workflow which we now briefly demonstrate by the construction of a target (\autorefFigExamples{}iii).

To start, we draw three concentric circles snapped to the same center point and change the color of the middle circle.
After selecting the three circles, we invoke \toolNameFirstUse{Repeat by Indexed Merge}, from which we select the second of two results (which differs from the first only in that it adds the \verb+reverse+ call seen below, so that \verb+i+ $=$ \verb+0+ for the smallest circle).

\begin{lstlisting}
...
circles =
  map (\i ->
      circle ??(1 => 0, 2 => 466, 3 => 0) \
             point \
             ??(1 => 114, 2 => 68, 3 => 25))
    (reverse (zeroTo 3{0-15}))    
...
\end{lstlisting}

The program maps an anonymous function that takes an index \verb+(\i -> ...)+ over the list \verb+[2,1,0]+; each index is thus transformed into one of the circles. The anonymous function is a syntactic merger of the original three circle definitions.
Their differences---radius and color---have been turned into what we call \emph{programming-by-example (PBE) holes}, represented by \verb+??(...)+. The first PBE hole above
can be read as ``the first time this expression is executed it should return \verb+0+, the second time it is executed it should return \verb+466+, and the third time \verb+0+''.

\begin{wrapfigure}{r}{0pt}
  \begin{minipage}{2in}
    \includegraphics[width=\textwidth]{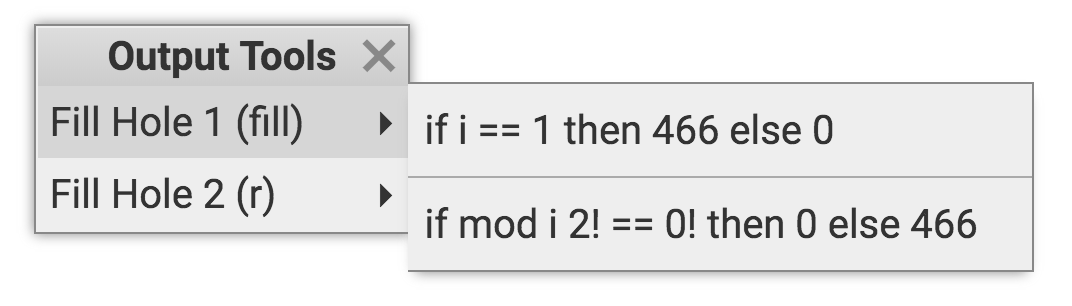}
    \includegraphics[width=\textwidth]{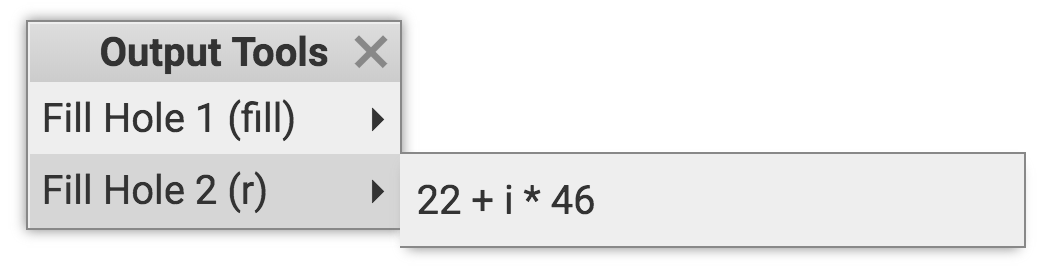}
  \end{minipage}
\end{wrapfigure}

\sns{} employs sketch-based synthesis~\cite{SketchingThesis} to resolve these holes.
As shown at right, all suggested fillings involve \verb+i+, the only variable
that differs in the execution environments.
For the first hole, we choose the \verb+mod i 2 == 0!+ conditional to obtain alternating colors. For the second we choose the only option, a $base \miniSepThree\mathtt{+}\miniSepThree i\miniSepTwo\mathtt{*}\miniSepTwo width$ expression, to calculate the radii of our circles. Similar to the Koch snowflake, the inserted code \verb+zeroTo 3{0-15}+ contains a slider annotation allowing us to manipulate the number of circles.
\autorefFigExamples{}xiii shows five circles for the final design.

\end{document}